\documentstyle[amsfonts,epsfig,12pt]{article}
\newcommand{\be}{\begin{equation}}
\newcommand{\ee}{\end{equation}}
\newcommand{\bea}{\begin{eqnarray}}
\newcommand{\eea}{\end{eqnarray}}
\newcommand{\beas}{\begin{eqnarray*}}
\newcommand{\eeas}{\end{eqnarray*}}
\newcommand{\bi}{\begin{itemize}}
\newcommand{\ei}{\end{itemize}}
\newcommand{\bc}{\begin{center}}
\newcommand{\ec}{\end{center}}
\newcommand{\bfl}{\begin{flushleft}}
\newcommand{\efl}{\end{flushleft}}
\newcommand{\bfr}{\begin{flushright}}
\newcommand{\efr}{\end{flushright}}

\def\6{\partial} \def\a{\alpha} \def\b{\beta}
\def\g{\gamma}

\def\m{\mu}

\def\o{\omega} \def\G{\Gamma}

\begin{document}
\title{Thermal D-branes States from Superstrings in Light-Cone Gauge }

\author{Ion V. Vancea\footnote{ionvancea@ufrrj.br}\\
         %\thanks{A footnote may follow.}\\
        Departamento de F\'{\i}sica Te\'{o}rica,\\
        Universidade Federal Rural do Rio de Janeiro (UFRRJ), Brazil\\
        BR 465-07-Serop\'{e}dica CEP 23890-000 RJ}
\maketitle

%\author{Another Author\\
%        Affiliation\\
%        E-mail: \email{...}}

\begin{abstract}In this talk we are going to review a method to construct the thermal boundary states of the thermal string in the TFD approach. The class of thermal boundary states presented here is derived from the BPS D-branes of the type II GS superstrings. 
\end{abstract}  

\newpage

\section{Introduction}

The representation of the $D$-branes as boundary states of the perturbative string is a powerful technique in the investigation of the geometrical and physical properties of the $D$-branes. In particular, it has been shown recently in \cite{ivv1,ivv2,ivv3,ivv4,ivv5,ivv6} that this framework can be used to give a microscopic description of the statistical properties of the bosonic $D$-branes. Since the $D$-branes are constructed out of states from the Fock space of the closed bosonic string, it is natural to approach the problem within the Thermo Field Dynamics (TFD) formalism which can be formulated in the canonical quantization \cite{tubook}. The TFD was applied previously to the string theory, e. g. in \cite{yl1,fns1}. More recent results were obtained in \cite{ivv7,ng1,ng2}. It deserves to emphasize two recent results from \cite{ng3} where the equivalence between the Matsubara and the TFD formalisms was proved and \cite{ng4} where the extension of the formalism to the superstring in the $pp$-background was done. (For more references on the application of the TFD to strings and $D$-branes see \cite{ivv8} and the references therein.) The aim of this paper is to report on the generalization of the method from \cite{ivv1} to construct the thermal $D$-branes obtained from the BPS $D$-branes of the GS type II superstring. The $D$-branes of the GS superstring at zero temperature were obtained for the first time in \cite{mg,gg}. As the RNS $D$-branes, they display the structure of coherent boundary states, and thus their are suited to the TFD analysis.

The structure of the paper is as follows. In Section 2 we are going to review the construction of the BPS $D$-branes of the type II superstring. In Section 3 we apply the TFD to the superstring and obtain the thermal string. In Section 4 we construct the thermal boundary states of the thermal string. The Section 5 is devoted to discussions. The present report reviews the results communicated in \cite{ivv8}.  

\section{Type II $D$-Branes in the Light-Cone Gauge} 

The main advantage of the GS formulation of the type II superstrings is that it displays the space-time supersymmetry at both classical and quantum level. However, the quantization is simple only in the light-cone gauge in which it coincides with the RNS quantization. The light-cone gauge action is given by the following relation \cite{gsw}
\be
I_{l.c.g.} = -\frac{1}{2} \int d^2 \sigma \left(T \6_{\a}X^{I}\6^{\a}X^{I} - \frac{i}{\pi}\bar{S}^{*A} \Gamma^{-} \rho^{\a}_{AB}\6_{\a}S^{B}\right),
\label{stringaction}
\ee
where $S^{*Aa}=S^{\dagger Bb}(\G^{0})^{ba}(\rho^0)^{BA}$ represents the conjugation of a spinor in two and eight dimensions and $A, B = 1,2$. By $\rho^{\a}$'s we denote the complex Dirac matrices in two dimensions while $\G^{\m}$'s are the $32 \times 32$ imaginary Dirac matrices of the spinor representation of $SO(1,9)$ and $\g^{I}$'s stand for the $16 \times 16$ real symmetric Dirac matrices of the spinor representation of $SO(8)$.

The action (\ref{stringaction}) has two supersymmetries. In the type IIA superstring, the supercharges that generate the supersymmetries can be written as two inequivalent spinors $Q^{a}$ and $Q^{\dot{a}}$ from $\mathbf{{8}_{s}}$ and $\mathbf{{8}_{c}}$, respectively, given by the following relations
\be
Q^{a} = \frac{1}{\sqrt{2p^{+}}}\int^{\sigma}_{0} d\sigma \, S^{a}(\sigma )~,~
Q^{\dot{a}} = \frac{1}{\pi\sqrt{p^{+}}}\int^{\sigma}_{0} d\sigma \, \gamma^{I}_{\dot{a}b} \6 X^{I}  S^{b}(\sigma ).
\label{charges}
\ee
Similar relations hold for the right-moving supercharges. In the type IIB theory the two supercharges belong to the same representation.

The BPS $Dp$-branes are defined by imposing $I=1,2,\ldots , p+1$ Neumann boundary conditions and $I=p+2,\ldots , 8$ Dirichlet boundary conditions on the superstring equations of motion and requiring that half of the supersymmetries of the theory be preserved. For definites, let us choose the type IIB superstring theory and define the following linear combinations of supercharges
\be
Q^{\pm a}=(Q^{a}\pm iM_{ab}\bar{Q}^{b})~,~Q^{\pm \dot{a}}=(Q^{\dot{a}}\pm iM_{\dot{a}\dot{b}}\bar{Q}^{\dot{b}}).
\label{lincharge}
\ee
By quantization, the charges $Q^{\pm a}$ and $Q^{\pm \dot{a}}$ become operators that act on the Hilbert space of the superstring. The breaking of half of supersymmetries can be written as constraints on the Hilbert space \cite{gg}
\bea
(\partial X^{I}-M_{J}^{I}\overline{\partial}X^{J})|B\rangle=0\nonumber,\\
Q^{+a}|B\rangle=Q^{+\dot{a}}|B\rangle=0\label{zeroTbc}.
\eea
The matrices $M^{I}_{J}$, $M_{ab}$ and $M_{\dot{a}\dot{b}}$ depend on the background fields and on the boundary conditions on the bosonic fields.

As shown for the first time in \cite{mg}, the type II superstring has $D$-branes boundary states in the light-cone gauge of the form
\cite{gg}
\bea
|B\rangle &=& 
\exp \sum_{n>0}\left( M_{IJ}a^{I\dagger}_{n}\bar{a}^{J\dagger}_{n} - iM_{ab}S^{a\dagger}_{n}\bar{S}^{b\dagger}_{n}     \right)|B_0\rangle,\label{zeroTDbrane}\\
|B_0\rangle &=& (M_{IJ}|I\rangle|J\rangle +iM_{\dot{a}\dot{b}}|\dot{a}\rangle |\dot{b}\rangle)\label{zeroTDbranezero}.
\eea
Here, we have used the Fourier expansion of the superstring fields in the normalized form. 
The superstring states from (\ref{zeroTDbranezero}) are obtained by acting with the bosonic and fermionic creation operators on the degenerate massless ground state.

\section{Thermalization of the GS Superstring}

The type II superstring can be heated by putting it in contact with a thermal reservoir. According to the TFD prescription \cite{tubook} each string oscillator will interact with the corresponding degree of freedom of the reservoir and the result is the thermalized string.  The particularities of superstring theory in the TFD framework are: the degenerate massless background at $T = 0$ and the constraints from the Virasoro algebra. Therefore, the TFD ansatz \cite{tubook} should be modified to \cite{ng2,ivv8}
\be
\langle O \rangle = Z(\b_T )^{-1} \mbox{Tr} \left[ \delta(P = 0) e^{-\b_T H } O\right] \equiv
\langle\langle 0(\b_T )| O |0(\b_T )\rangle\rangle,
\label{partfunct}
\ee 
for any observable $O$. In the above relation the delta function implements the level matching condition in the trace over the Hilbert space of the string theory in order to pick up just the contributions from the physical subspace. The ansatz (\ref{partfunct}) defines the thermal vacuum $|0(\b_T )\rangle\rangle$ as a vector from the tensor product of the superstring physical Hilbert space with an identical copy of it denoted by $\tilde{}$. Its explicit form is given by the following relation
\be
|0(\b_T )\rangle\rangle = e^{-iG}|0\rangle\rangle~,~|0\rangle\rangle = |0\rangle|\tilde{0}\rangle,
\label{thermvacuum}
\ee
where $G$ is the Bogoliubov operator of all superstring oscillator modes \cite{tubook}
\be
G^B = \sum_{n=1}^{\infty}\, (G^{B}_n + \bar{G}^{B}_n )~,~
G^F = \sum_{n=1}^{\infty}\, (G^{F}_n + \bar{G}^{F}_n ),
\label{BogOpn}
\ee  
and $G^B$ and $G^F$ denote the sums of bosonic and fermionic Bogoliubov operators, respectively,
\bea
G^{B}_n &=& -i\theta^{B}_{n}(\beta_T)\sum_{I=1}^{8}(a^{I}_n \, \tilde{a}^{I}_n - \tilde{a}^{I\dagger}_n \, a^{I \dagger}_n )\label{GBn},\\
\bar{G}^{B}_{n} &=& -i\theta^{B}_{n}(\beta_T)\sum_{I=1}^{8}(\bar{a}^{I}_n \, \tilde{\bar{a}}^{I}_n - \tilde{\bar{a}}^{I\dagger}_n \, \bar{a}^{I \dagger}_n )\label{barGBn},\\
G^{F}_{n} &=& -i\theta^{F}_{n}(\beta_T)\sum_{a=1}^{8}(S^{a}_n \, \tilde{S}^{a}_n - \tilde{S}^{a\dagger}_n \, 
S^{a \dagger}_n )\label{GFn},\\
\bar{G}^B &=& -i\theta^{F}_{n}(\beta_T)\sum_{a=1}^{8}(\bar{S}^{a}_n \, \tilde{\bar{S}}^{a}_n - \tilde{\bar{S}}^{a\dagger}_n \, \bar{S}^{a \dagger}_n )
\label{barGFn}.
\eea
There is no TFD prescription for the thermalization of zero mode operators from the fermionic sectors. However,  $S^{a}_{0}$, $\bar{S}^{a}_{0}$, $\tilde{S}^{a}_{0}$ and $\tilde{\bar{S}}^{a}_{0}$ are isomorphic to the Dirac matrices. Therefore, we consider them inert under the thermalization. Note that these operators are important for constructing the thermal states from the thermal vacuum in analogy with zero temperature case. In contrast with the $T=0$ situation, the thermal vacuum does not have zero mass. Indeed, since the mass is an eigenvalue of zero temperature mass operator {\em not} finite temperature mass operator, the Bogoliubov operator creates massive states from the zero temperature vacuum.

An useful form for calculations of the Bogoliubov operator of a single oscillator is the  linear form. In the case of superstrings, the Bogoliubov transformations of the superstring oscillators act linearly on the oscillator operators as follows
\bea
a^{I}_n & \longrightarrow & a^{I}_n(\b_T ) = a^{I}_n \cosh \theta^{B,I}_n(\b_T ) - \tilde{a}^{I\dagger}_n \sinh \theta^{B,I}_n(\b_T ),\label{tra}\\
a^{I \dagger}_n & \longrightarrow & a^{I\dagger }_n(\b_T ) = a^{I \dagger}_n \cosh \theta^{B,I}_n(\b_T ) - \tilde{a}^{I}_n \sinh \theta^{B,I}_n(\b_T ),\label{tradagger}\\
\tilde{a}^{I}_n & \longrightarrow & \tilde{a}^{I}_n(\b_T ) = \tilde{a}^{I}_n \cosh \theta^{B,I}_n(\b_T ) - a^{I\dagger}_n \sinh \theta^{B,I}_n(\b_T ),\label{trtildea}\\
\tilde{a}^{I\dagger}_n & \longrightarrow & \tilde{a}^{I \dagger}_n(\b_T ) = \tilde{a}^{I \dagger}_n \cosh \theta^{B,I}_n(\b_T ) - a^{I}_n \sinh \theta^{B,I}_n(\b ),\label{trtildeadagger}\\
\bar{a}^{I}_n & \longrightarrow & \bar{a}^{I}_n(\b_T ) = \bar{a}^{I}_n \cosh \bar{\theta}^{B,I}_n(\b_T ) - \tilde{\bar{a}}^{I\dagger}_n \sinh \bar{\theta}^{B,I}_n(\b_T ),\label{bartra}\\
\bar{a}^{I \dagger}_n & \longrightarrow & \bar{a}^{I\dagger }_n(\b_T ) = \bar{a}^{I \dagger}_n \cosh \bar{\theta}^{B,I}_n(\b_T ) - \tilde{\bar{a}}^{I}_n \sinh \bar{\theta}^{B,I}_n(\b_T ),\label{bartradagger}\\
\tilde{\bar{a}}^{I}_n & \longrightarrow & \tilde{\bar{a}}^{I}_n(\b_T ) = \tilde{\bar{a}}^{I}_n \cosh \bar{\theta}^{B,I}_n(\b_T ) - \bar{a}^{I\dagger}_n \sinh \bar{\theta}^{B,I}_n(\b_T ),\label{bartrtildea}\\
\tilde{\bar{a}}^{I\dagger}_n & \longrightarrow & \tilde{\bar{a}}^{I \dagger}_n(\b_T ) = \tilde{a}^{I \dagger}_n \cosh \bar{\theta}^{B,I}_n(\b_T ) - \bar{a}^{I}_n \sinh \bar{\theta}^{B,I}_n(\b_T ),\label{bartrtildeadagger}\\
S^{a}_n &\longrightarrow & S^{a}_n (\b_T) = S^{a}_n \cos \theta^{F,a}_{n}(\b_T ) - \tilde{S}^{a\dagger}_n \sin \theta^{F,a}_{n}(\b_T ),\label{trS}\\
S^{a \dagger}_n &\longrightarrow & S^{a \dagger}_n (\b_T) = S^{a \dagger}_n \cos \theta^{F,a}_{n}(\b_T ) - \tilde{S}^{a}_n \sin \theta^{F,a}_{n}(\b_T ),\label{trSdagger}\\
\tilde{S}^{a}_n &\longrightarrow & \tilde{S}^{a}_n (\b_T) = \tilde{S}^{a}_n \cos \theta^{F,a}_{n}(\b_T ) + S^{a\dagger}_n \sin \theta^{F,a}_{n}(\b_T ),\label{trtildeS}\\
\tilde{S}^{a\dagger}_n &\longrightarrow & \tilde{S}^{a \dagger}_n (\b_T) = \tilde{S}^{a \dagger}_n \cos \theta^{F,a}_{n}(\b_T ) + \tilde{S}^{a}_n \sin \theta^{F,a}_{n}(\b_T ),\label{trtildeSdagger}\\
\bar{S}^{a}_n &\longrightarrow & \bar{S}^{a}_n (\b_T) = \bar{S}^{a}_n \cos \bar{\theta}^{F,a}_{n}(\b_T ) - \tilde{\bar{S}}^{a\dagger}_n \sin \bar{\theta}^{F,a}_{n}(\b_T ),\label{bartrS}\\
\bar{S}^{a \dagger}_n &\longrightarrow & \bar{S}^{a \dagger}_n (\b_T) = \bar{S}^{a \dagger}_n \cos \bar{\theta}^{F,a}_{n}(\b_T ) - \tilde{\bar{S}}^{a}_n \sin \bar{\theta}^{F,a}_{n}(\b_T ),\label{bartrSdagger}\\
\tilde{\bar{S}}^{a}_n &\longrightarrow & \tilde{\bar{S}}^{a}_n (\b_T) = \tilde{S}^{a}_n \cos \bar{\theta}^{F,a}_{n}(\b_T ) + \bar{S}^{a\dagger}_n \sin \bar{\theta}^{F,a}_{n}(\b_T ),\label{bartrtildeS}\\
\tilde{\bar{S}}^{a\dagger}_n &\longrightarrow & \tilde{\bar{S}}^{a \dagger}_n (\b_T) = \tilde{\bar{S}}^{a \dagger}_n \cos \bar{\theta}^{F,a}_{n}(\b_T ) + \tilde{\bar{S}}^{a}_n \sin \bar{\theta}^{F,a}_{n}(\b_T ).\label{bartrtildeSdagger}
\eea
Here, $\theta^{B,I}_{n}(\beta_T)$ and $\theta^{F,a}_{n}(\beta_T)$ are factors that depend on the temperature through $\beta_T = (k_B T)^{-1}$ and on the $n$-th oscillator frequency $\o_n$. Therefore, these factors will differ only for different oscillators $n$, but once $n$ is given, they are the same in all spacetime directions $I=1,\ldots ,8$ and for all spinor components $a=1,\ldots ,8$, respectively. Thus, the $I$ and $a$ indices can be dropped. Also, since the left-moving and right-moving oscillators of the same $n$ should be identical by symmetry between the left-moving and right-moving modes of the closed superstring, the $\theta_n$-factors should be the same in both left-moving and right-moving sectors and $\ \bar{} \ $ can be dropped, too. The form of $\theta$'s is given by the following relations \cite{tubook}
\be
\theta^{B}_{n}(\beta_T ) = \mbox{arccosh} (1-e^{-\beta_T \omega^{B}_n})^{-\frac{1}{2}}~,~
\theta^{F}_{n}(\beta_T ) = \mbox{arccos} (1+e^{-\beta_T \omega^{F}_n})^{-\frac{1}{2}},
\label{thetas}
\ee
where $\omega^{B}_n$ and $\omega^{F}_{n}$ are the frequencies of $n$-th bosonic and fermionic oscillators, respectively.

\section{Thermal $D$-brane States}

The similarity transformations generated by the Bogoliubov operator can be applied to the operators that define the $D$-brane boundary conditions (\ref{zeroTbc}) and the following boundary conditions at $ T \neq 0$ are obtained
\bea
(\partial X^{I}(\b_T)-M_{J}^{I}\overline{\partial}X^{J}(\b_T))|B(\b_T)\rangle\rangle=0,
\label{Tbctotalstring1}\\
Q^{+a}(\b_T)|B(\b_T)\rangle \rangle = Q^{+\dot{a}}(\b_T)|B(\b_T)\rangle\rangle=0\label{Tbctotalstring2},\\
(\partial \tilde{X}^{I}(\b_T)-M_{J}^{I}\overline{\partial}\tilde{X}^{J}(\b_T))|B(\b_T)\rangle\rangle=0,
\label{Tbctotaltildestring1}\\
\tilde{Q}^{-a}(\b_T)|B(\b_T)\rangle \rangle = \tilde{Q}^{-\dot{a}}(\b_T)|B(\b_T)\rangle\rangle=0 .\label{Tbctotaltildestring2}
\eea
The above equations can be solved by noting that $ T =0 $ boundary states can be factorized in two factors belonging to the superstring and the tilde-superstring Hilbert spaces, respectively (for details see \cite{ivv8}.) The thermalized $D$-brane states have the following form   
\be  
|B(\b_T)\rangle\rangle = e^{\Sigma(\b_T) + \tilde{\Sigma}(\b_T)}|B_0(\b_T)\rangle\rangle,
\label{thermalsol1}
\ee
where we have introduced the following notations
\bea
\Sigma(\b_T) &=& \sum_{n>0}( M_{IJ}a^{I\dagger}_{n}(\b_T)\bar{a}^{J\dagger}_{n}(\b_T) - iM_{ab}S^{a\dagger}_{n}(\b_T)\bar{S}^{b\dagger}_{n}(\b_T)), \label{sigmaop}\\    
\tilde{\Sigma}(\b_T) &=& \sum_{n>0}( \tilde{M}_{IJ}\tilde{a}^{I\dagger}_{n}(\b_T)\tilde{\bar{a}}^{J\dagger}_{n}(\b_T) + i\tilde{M}_{ab}\tilde{S}^{a\dagger}_{n}(\b_T)\tilde{\bar{S}}^{b\dagger}_{n}(\b_T)),\label{tildesigmaop}\\
|B_0(\b_T)\rangle\rangle &=& (M_{IJ}|IJ(\b_T)\rangle\rangle +iM_{\dot{a}\dot{b}}|\dot{a}\dot{b}(\b_T)\rangle\rangle)
(M_{IJ}|\tilde{I}\tilde{J}(\b_T)\rangle\rangle -iM_{\dot{a}\dot{b}}|\tilde{\dot{a}}\tilde{\dot{b}}(\b_T)\rangle\rangle).
\label{thermalzeromode3}
\eea
The states $|IJ(\b_T)\rangle\rangle$ are obtained by acting with $a^{I\dagger}_{1}(\b_T)\bar{a}^{J\dagger}_{1}(\b_T)$ on the thermal vacuum $|0(\b_T)\rangle\rangle$, in analogy with zero temperature.

\section{Discussions}

In the present communication we have reviewed the construction of the thermal boundary states from the BPS $D$-branes of the type II superstring in GS formulation by thermalization in the TFD framework. The boundary relations at finite temperature (\ref{Tbctotalstring1}), (\ref{Tbctotalstring2}), (\ref{Tbctotaltildestring1}) and (\ref{Tbctotaltildestring2}) can be obtained from the thermalized action of the total string, i. e. the difference between the action of the superstring and the tilde-superstring. However, by thermalization most of the superstring symmetries are broken. In particular, as argued in \cite{ivv8} the $\epsilon$-supersymmetry is broken while the $\eta$-supersymmetry is preseved. This pattern is preserved on the full thermal Hilbert space. Therefore, the thermal boundary state (\ref{thermalsol1}) is not supersymmetric. 

The formalism presented here could be used to compute the entropy of the thermal string in the presence of the $D$-brane, which was obtained in \cite{vm} by the Matsubara method. Also, it could be used to study the thermal string in curved backgrounds, too, at least in some approximation in which there is a canonical quantization of the theory. We hope to report on this topic soon \cite{hemi}. 

{\bf Acknowledgments}
I would like to thank to the organizers of the \emph{Fifth International Conference on Mathematical Methods in Physics} for the invitation.

\end{document}